\documentclass[oneside,letterpaper]{amsart}
\usepackage[latin1]{inputenc}
\usepackage{amssymb}
\usepackage{amsrefs}
\usepackage{geometry}
\usepackage{float}
\usepackage{graphicx}

\usepackage{appendix}

\makeatletter
\theoremstyle{plain}
\newtheorem{thm}{Theorem}[section]
\theoremstyle{plain}

\theoremstyle{remark}
\newtheorem*{acknowledgement*}{Acknowledgement}
\theoremstyle{plain}
\theoremstyle{definition}
\newtheorem*{defn*}{Definition}
\newtheorem{defn}{Definition}

\makeatother

\begin{document}

\title{A Gauge-independent Mechanism for Confinement and Mass Gap: Part I --- The General Framework}

\author{J. Wade Cherrington$^{1}$}

\maketitle
\vspace*{-10pt}

\begin{center}{\small $^{1}$Department of Mathematics, University
of Western Ontario, London, Ontario, Canada \newline }\par\end{center}{\small \par}

\begin{abstract}
We propose a gauge-independent mechanism for the area-law behavior of Wilson
loop expectation values in terms of worldsheets spanning Wilson loops interacting with
the spin foams that contribute to the vacuum partition function.
The method uses an exact transformation of lattice-regularized Yang-Mills theory that is valid
for all couplings. Within this framework, some natural conjectures can be
made as to what physical mechanism enforces the confinement property in the continuum (weak coupling)
limit. Details for the $SU(2)$ case in three dimensions are provided in a companion paper.
\end{abstract}

\section{Background}
Our current understanding of confinement in Yang-Mills gauge theories encompasses a seemingly diverse collection
of approaches, including center vortices~\cite{cvortexthooft}, dual superconductors~\cite{dsuperthooft,
dsupermandelstam,dsuperparisi}, and effective strings of various types~\cite{stringynambu, stringyantonov} ---
to name just a few that have attracted interest over the years.

In the present work we propose a mechanism for confinement that is similar in concept to the well-known 
proof of confinement in the (unphysical) strong coupling limit~\cite{M1980}, but suitably generalized 
so that the analysis is valid at the arbitrarily weak couplings that are characteristic of the continuum
limit. Our main result is a new form of the Wilson loop criteria for confinement in lattice gauge theory, 
in terms of which we conjecture a general confinement scenario that can be applied to any specific Yang-Mills 
theory (i.e. choice of gauge group and space-time dimensionality).

We refer the reader to a companion paper~\cite{Conf2} for a demonstration of the 
framework developed herein for the case of $G=SU(2)$ in three dimensions.
The present paper focuses on establishing the worldsheet-background decomposition of spin foams and
outlining in general terms those aspects of our proposal that are model independent. For 
concreteness, we do however specialize to the case of $SU(2)$ starting in Section 3. The scenario 
is also well-defined in the $SU(3)$ case, but more detailed analysis will require further developments in the 
evaluation and asymptotic limits of $SU(3)$ spin networks.

The structure of the present paper is as follows. In Section 2, we shall review the standard lattice 
regularization of pure Yang-Mills theory on the lattice, the definition of a Wilson loop observable, and 
the concept of character expansion for the conventional amplitude. This will allow us to give a brief account
of the widely known proof of confinement at strong coupling and fix basic notation to be used throughout.

In Section 3, we will introduce the spin foam formulation of lattice Yang-Mills theory, which is exactly dual 
to the conventional formulation. By analysis of admissibility conditions, we show that contributions to the 
Wilson loop expectation value can be organized as a double sum over worldsheets connected to the Wilson loop and 
over vacuum spin foams, a result already implicit but only realized in a limited way in past work~\cite{IDV2} on 
strong coupling.  In Section 4 we show how this decomposition of spin foam configurations can lead to a natural 
generalization of the strong coupling analysis of confinement. We also discuss how certain features of a 
roughening transition are present in transitioning to the weak coupling limit but present no obstacle within 
the spin foam framework, unlike past methods that use strong coupling expansions.
We conclude in Section 5 with a discussion of our results and the further lines of inquiry that they suggest.

\section{Review of lattice formulation and confinement at strong coupling}
Recall the Euclidean partition function of pure (non-abelian) Yang-Mills theory in $D$ dimensions,
with gauge group $G=SU(N)$ for $N\geq2$. In the continuum it takes the form 
\begin{equation}
        Z_{V}=\int\mathcal{D}A\,\exp(-S [A] ),\label{eq:YMpartfunc}
\end{equation}
where $A = A_{\mu}^{a}(x)$ is the gauge field, $S$ is the action functional, and $\mathcal{D}A$ is
the functional integration measure.
\begin{defn}[Lattice]
We shall denote our oriented $D-$dimensional hypercubic lattice by $\kappa$, and its associated set of 
vertices, edges, and plaquettes by $V,$ $E$, and $P$, respectively.
\end{defn}
We apply the standard lattice regularization by defining a functional integral measure as an 
integral over a product of $|E|$ copies of $G$ using the Haar measure: 
\begin{equation}
        \mathcal{D}A\equiv \prod_{e\in E}dg_{e}.
\end{equation}
To proceed we replace the continuum action functional by a discretized
action, $S\equiv \hat{S}[g]$ that reproduces the continuum action $S[A]$
in the limit of zero lattice spacing. A discretized action functional $\hat{S}$ can
be formed by summing over a local function $S(g_{p})$ of variables at a plaquette,
$\hat{S}[g]=\sum_{p\in P}S(g_{p})$, where
the group element $g_{p}$ is the holonomy around an oriented fundamental plaquette
$p$. That is, $g_{p}=g_{1}g_{2}g_{3}g_{4}$, where $g_{i}$
is either the group element assigned to the $i$th edge of $p$ or
its inverse if the orientations of $p$ and the $i$th edge are opposing.
This yields the conventional lattice partition function
\begin{equation}
        Z_{V}=\int \left( \prod_{e\in E}dg_{e} \right) e^{-\sum_{p\in P}S(g_{p})}.
\label{eq:conventional}\end{equation}
Let $\Gamma$ denote a lattice loop (a closed, non-self-intersecting
lattice curve). We define a \emph{Wilson loop} to
be a pair $(\Gamma,w)$ where $\Gamma$ is a lattice loop and $w$
is the label of the irreducible representation in which the trace
is to be taken. For brevity we denote this data by $\Gamma_{w}\equiv(\Gamma,w)$;
the object $\Gamma_{w}$ can be thought of as a loop in the lattice
colored by the charge $w$. The expectation value $\left\langle O_{\Gamma_{w}}\right\rangle$ associated
with $\Gamma_{w}$ is defined as a field insertion into the partition function as follows:
\begin{equation}
\left\langle O_{\Gamma_{w}}\right\rangle \equiv \frac{\int \left( \prod_{e\in E}dg_{e} \right) \,
\text{Tr} \left( \prod_{e\in\Gamma}D_{w}(g_{e}) \right) e^{-\sum_{p\in P}S(g_{p})}}
{\int \left( \prod_{e\in E}dg_{e} \right)\, e^{-\sum_{p\in P}S(g_{p})}}.
\label{eq:WilsonExpect1}
\end{equation}
Here $\prod_{e\in\Gamma}D_{w}(g_{e})$ denotes a product of representation matrices $D_{w}(g)$
in the $w$th irreducible representation of $G$ (in studying confinement for $G=SU(2)$, the
case $w=\frac{1}{2}$ is usually considered).
We can now give a brief sketch of the proof of confinement (at strong coupling) by way of the 
\emph{character expansion}. This approach to expanding the partition function is an exact 
transformation, as the summation over all diagrams yields a convergent sum for any non-zero value
of the coupling constant~\cite{M1980}. We have for the vacuum partition function the following expansion:
\begin{eqnarray}
Z_{V}&=&\int \left( \prod_{e\in E}dg_{e} \right) \prod_{p\in P}\left(\sum_{j_{p}=\frac{1}{2}}^{\infty}
c_{j_{p}}\chi_{j_{p}}(g_{p})\right) \\
&=&\sum_{\{ j_{p}\}}^{\infty}\int \left( \prod_{e\in E}dg_{e} \right) \prod_{p\in P}{c_{j_{p}}\chi}_{j_{p}}(g_{p}),
\nonumber
\end{eqnarray}
where $e^{-S(g_{p})}=\sum_{j_{p}=\frac{1}{2}}^{\infty}c_{j_{p}}\chi_{j_{p}}(g_{p})$ defines
the character coefficients $c_{j_{p}}$ associated with the chosen plaquette action $S(g_p)$.
In the second line we have interchanged the order of summation and integration operations. 
The notation $\{ j_{p}\}$ refers to the set of all labellings of the plaquettes of $\kappa$ by unitary 
irreducible representations (irreps) of $G$. Each such labelling can be viewed as a
``diagram'' --- however, only a special subset of diagrams in the summation are non-zero.
The labellings\footnote{Note we are not following the more traditional method of expanding
in powers of $\beta$, but rather products of character coefficients
(see for example \cite{DZ1983} for more on this approach to strong coupling
expansions). } which give non-zero (we refer to these as \emph{admissible}) contributions to
$Z_{V}$ satisfy certain constraints on the representations of plaquettes that meet at an edge. 
In $SU(2)$, a geometric characterization is possible: admissible diagrams can be viewed as 
collections of surfaces that are closed (no free edges allowed), branched (can meet other surfaces
along vertices, edges, and plaquettes) and colored by the group irrep
labels~\cite{DZ1983, MandM, ConradyKhavkine}.

By applying the same procedure of character expansion described for the vacuum partition function 
above, one can associate a charged partition
function $Z_{\Gamma}$ with the numerator of the expectation value~(\ref{eq:WilsonExpect1}):
\begin{eqnarray}\label{eqn:Z_gamma}
Z_{\Gamma} &\equiv& \int \left( \prod_{e\in E}dg_{e} \right)\,
\,\text{Tr} \left( \prod_{e\in\Gamma}D_{\frac{1}{2}}(g_{e}) \right) e^{-\sum_{p\in P}S(g_{p})} \\ \nonumber
&=&\sum_{\{ j_{p}\}}^{\infty}\int \left( \prod_{e\in E}dg_{e} \right) \text{\, Tr}\left
(\prod_{e\in\Gamma}D_{\frac{1}{2}}(g_{e})\right)\prod_{p\in P}c_{j_{p}}\chi_{j_{p}}(g_{p}).
\end{eqnarray}
As a function of Wilson loop $\Gamma$, the numerator of the expectation value $\left\langle O_{\Gamma_{w}}\right\rangle$
can thus be re-expressed as a sum over diagrams labelled by the unitary irreducible 
representations of $G$ as:
\begin{eqnarray}\label{eqn:wilsonexpectation}
\left\langle O_{\Gamma_{w}}\right\rangle = \frac{Z_{\Gamma}}{Z_{V}} = Z_{V}^{-1}\sum_{\{ j_{f}\}}^{\infty}
\int \left( \prod_{e\in E}dg_{e} \right)\,\text{Tr}\left(\prod_{e\in\gamma}
D_{\frac{1}{2}}(g_{e})\right)\prod_{p\in P}c_{j_{p}}\chi_{j_{p}}(g_{p}).
\end{eqnarray}
The summation is again over all labellings of the plaquettes of the lattice by unitary
irreps of $G$. However, due to the presence of representation matrices
associated with the Wilson loop, the admissibility constraints are locally modified along $\Gamma$. 
One important consequence of this is that for any contribution to the numerator of 
$\left\langle O_{\Gamma_{w}}\right\rangle$ there is a lower bound on the number of plaquettes
carrying non-trivial irrep labels that is proportional to the area of a minimal surface spanning the loop. As we shall argue 
more rigorously below, the expectation value of a Wilson loop operator can be expressed as a ratio
of two partition functions, involving two different collections of surfaces: one in which all surfaces
close ($Z_V$) and a second in which a surface always ends on the Wilson loop, while all other surfaces
close ($Z_{\Gamma}$).

Transition to the continuum limit is controlled by $\beta$, a parameter inversely proportional to the
coupling constant.  Although suppressed in the above notation, $\beta$ enters through the character
coefficients $c_{j_{p}}=c_{j_{p}}(\beta)$. Later, we shall see how the character coefficients become plaquette amplitudes in the
spin foam formulation, and the fact that they contain all the $\beta$-dependence of the model will be important
to understanding the weak coupling limit. 
Returning now to~(\ref{eqn:wilsonexpectation}), it can be shown that for sufficiently small $\beta$, the
strong decay of $c_{j_{p}}(\beta)$ for large spin combined with bounds on the integrals of products of characters $\chi_{j}$
allows for a series expansion for $\left\langle O_{\Gamma_{w}}\right\rangle$ having a finite radius
of convergence and leading order terms that are exponentially decaying in the area of the Wilson loop~\cite{Osterwalder}.

As is typically remarked when considering the extension of these arguments to weak coupling, increasing $\beta$ 
(as required for the continuum limit) decreases the energetic cost associated with the area of the minimal 
spanning surface~\cite{IDV2}. As the number of worldsheets of a given area grows exponentially in that area, absent 
other effects, at a sufficiently large $\beta$ the entropy of surfaces will ``win out'', and the surface
will delocalize (the position of nearby plaquettes on the surface will become uncorrelated). 
This general phenomenon is referred to as a~\emph{roughening transition}~\cite{itzykrough}.  

In what follows, we show how to apply the spin foam form of dual lattice gauge theory to resolving each contribution
to $\left\langle O_{\Gamma_{w}}\right\rangle$ as a local interaction between a $\Gamma$-connected
fundamental worldsheet and a ``background'' spin foam from the vacuum ensemble.  Upon doing so it is clear 
from inspecting the amplitude (and verified by numerical simulations~\cite{CCK}), that increasing $\beta$
also leads to rapid growth in the average spin on the background spin foams that the worldsheet has to
interact with.
Analysis of this interaction will lead us to conjecture that it is this growth in the average background
spin and the resulting high energy cost of inserting large area worldsheets that provides confinement in the
weak coupling limit.

\section{Existence of Worldsheet-background decompositions for $SU(2)$}
In this section we will start by reviewing an equivalent formulation of Yang-Mills theory
on the lattice that uses spin foams, which are closely related to strong coupling diagrams
but allow the factorization of the amplitude into local degrees of freedom. 
The reader is referred to some of the existing literature for derivations and 
details on the spin foam formulation of lattice gauge theory~\cite{OecklPfeiffer,OecklDGT,LAT08} and 
earlier forms of non-abelian duality~\cite{AS, ACSM, AGMS91} in the specific
case of $D=3,G=SU(2)$.

Using the spin foam formulation, we will show how a Wilson loop observable can be expressed
as a sum over worldsheets bounded by the loop interacting with a sum over vacuum 
spin foam configurations. An explicit construction for this interaction energy can be given
and will turn out to be local to the worldsheet.

To begin, we define a spin foam for a $D$-dimensional lattice $\kappa$ and group $G$ as follows.
\begin{defn}[Lattice spin foam]
Given a group $G$ and a lattice $\kappa$ with edges $E$ and plaquettes $P$, a \emph{spin foam} is 
a pair of maps: a plaquette labelling $P \rightarrow \mathcal{R}(G)$ and an edge labelling
$E \rightarrow \mathcal{I}(G)$, where $\mathcal{R}(G)$ and $\mathcal{I}(G)$ are the unitary 
irreducible representations of $G$ and the intertwiners of $G$, respectively.
\end{defn}

Using the spin foam model dual to lattice Yang-Mills, the vacuum partition function can be written as
\begin{eqnarray}\label{eqn:vac_spinfoams}
Z_{V} = \sum_{f \in F_{V}} \mathcal{A}(f) = \sum_{f \in F_{V}} 
\prod_{p \in P} A_{P}(f(p)) \prod_{e \in E} A_{E}(f(e)) \prod_{v \in V} A_{V}(f(v)),
\end{eqnarray}
where $F_{V}$ denotes the set of lattice spin foams associated with the vacuum partition function. 
The amplitudes $A_{V}$, $A_{E}$, and $A_{P}$ are functions
that depend on the plaquette and edge variables that are local to each vertex, edge, and plaquette respectively. 
The set $f(v)$ consists of the irrep variables assigned to plaquettes incident to $v$ and to
the intertwiner variables on edges incident to $v$. The set $f(e)$ denotes the intertwiner 
variable assigned to $e$ and the irrep variables on plaquettes incident to $e$ by $f$.
The set $f(p)$ is simply the irrep value assigned by $f$ to $p$.
While the spin foam transformation may be carried out for any $SU(N)$, we shall restrict ourselves
to the case of $SU(2)$ in what follows, as methods for explicitly evaluating amplitudes and their asymptotics
are in this case more fully worked out at present than for $SU(3)$.

To characterize the topological structure of those spin foams that make a non-zero contribution to $Z_{V}$,
we introduce the notions of edge and spin foam admissibility for vacuum spin foams (where no external charge
is present).

\begin{figure}[h]
\includegraphics[scale=0.85]{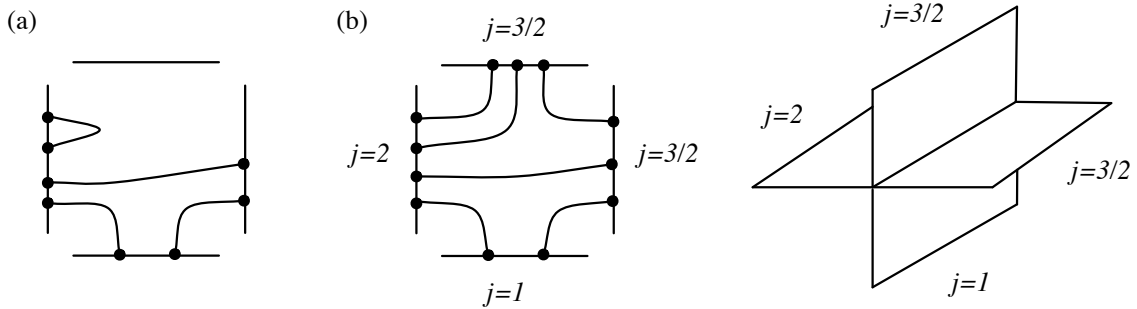}
\caption{(a) A crossingless matching with ``U-turn'' (does not correspond to any intertwiner) (b) Illustration of a crossingless 
matching corresponding to one possible intertwiner, given a labelling of plaquettes incident to an edge.}
\label{fig:matching}
\end{figure}

\begin{defn}[Edge vacuum admissibility]\label{def:edge_vacuum}
Let an edge $e \in E$ be given. Let $n$ equal the number of plaquettes incident to $e$, and $V_{j_{i}}$
the vector spaces associated to each representation label $j_{i}$.
An edge is \emph{admissible} if the space of intertwiners $I \equiv \text{Inv}(V_{j_{1}} \otimes V_{j_{2}} \cdots \otimes V_{j_{n}})$
is non-empty and the edge variable is labelled from the set $I$. A convenient graphical rule for
determining edge admissibility can be given as follows.

For every $V_{j_{i}}$, mark $\text{dim}(V_{j_{i}})-1$ points on the boundary of a disc, with each point associated with a given
irrep grouped beside each other. Then the dimension of the space 
$I \equiv \text{Inv}(V_{j_{1}} \otimes V_{j_{2}} \cdots \otimes V_{j_{n}})$ is equal to the number of crossingless
matchings in which no matching occurs between points in the same grouping, and each such crossingless matching\footnote{A 
\emph{matching} is a pairing of the boundary points of a disc by lines that connect them through the disc; the matching is \emph{crossingless} 
if the lines can be drawn in such a way that none of them intersect.} corresponds to a different intertwiner~\cite{K97}.  
An illustration of a crossingless matchings with and without U-turns is given in Figure~\ref{fig:matching} below (in this
example we work in $D=3$).
\end{defn}

\begin{defn}[Spin foam vacuum admissibility]
A spin foam is \emph{admissible} if $\forall e \in E$, edge admissibility is satisfied.
\end{defn}
We now turn to the modification of these conditions in the presence of a Wilson loop.
We seek a spin foam expansion of the numerator of equation~(\ref{eq:WilsonExpect1}), which we
shall define to be the externally charged partition function $Z_{\Gamma}$.
First we note that away from $\Gamma$, the spin foam amplitude is unchanged.
One consequence of this is the modification of vertex amplitudes along $\Gamma$. The spin foam amplitude
for the charged partition function is thus 
\begin{eqnarray}
\;\;\;\;\;\;\;\; Z_{\Gamma} = \sum_{f \in F_{\Gamma}} A_{\Gamma}(f)
\equiv \sum_{f \in F_{\Gamma}}  \nonumber
\prod_{p \in P} A_{P}(f(j))
\prod_{e \in E_{\Gamma}} \bar{A}_{E}(f(e))
\prod_{v \in V_{\Gamma}} \bar{A}_{V}(f(v)).
\end{eqnarray}
where
\begin{eqnarray}
&\bar{A}_{V}(\{j\}_v,\{i\}_v)& = 
\begin{cases} A_{V}(f(v)) \quad \text{for $v \notin \Gamma$} \\
              A^{*}_{V}(f(v),\Gamma) \quad \text{for $v \in \Gamma$}
\end{cases} \\ \nonumber
&\bar{A}_{E}(\{j\}_{e},i_e)& = 
\begin{cases} A_{E}(f(e)) \quad \text{for $e \notin \Gamma$} \\
              A^{*}_{E}(f(e)) \quad \text{for $e \in \Gamma$}
\end{cases}
\end{eqnarray}
For edges and vertices that are part of $\Gamma$, modified
amplitudes $A^{*}_{E}$ and $A^{*}_{V}$ are necessary (see for example~\cite{CherringtonFermions,ConradyPolyakov} for
details) and in the case of $A^{*}(V)$ depend on whether $\Gamma$ passes straight through $v$ or makes a
turn.

As with the vacuum diagrams, we seek to characterize the non-zero contributions to $Z_{\Gamma}$ in terms
of local admissibility conditions.  As the insertion of the trace around $\Gamma$ only introduces
representation matrices depending on $g_e$ for $e \in \Gamma$, admissibility conditions on edges
away from $\Gamma$ are unaffected.  Thus for defining admissibility with regard to $Z_{\Gamma}$ it
suffices to consider $e \in \Gamma$, as follows:
\begin{defn}[Edge $\Gamma$-admissibility]\label{def:edge_charge}
Let an edge $\epsilon$ belonging to a Wilson loop $\Gamma$ be given, let $n$ be equal to the number of plaquettes 
incident to $\epsilon$, and let $V_{j_{i}}$ denote the vector spaces associated to each plaquette irrep 
label $j_{i}$. We say that $\epsilon$ is \emph{$\Gamma$-edge admissible} if and only if the space 
$\text{Inv}(V_{\frac{1}{2}} \otimes V_{j_{1}} \otimes V_{j_{2}} \cdots \otimes V_{j_{n}})$ is
non-empty and the edge is labelled by an intertwiner belonging to this non-empty set.

This leads to a charged edge admissibility condition corresponding to a crossingless matching structure with
the incident representations appearing as in the vacuum case, plus a single additional point, which we shall 
call the \emph{charged point}. The condition is illustrated graphically in Figure~\ref{fig:edgecharge} below.
\end{defn}

We note in passing that this definition can be applied to a loop $\Gamma$ with multiple
connected components; e.g. a Wilson loop observable with two disconnected loops separated by
some distance, such as is used in computing the two-point expectation value
(See Appendix~\ref{append:massgap}). 

Since the group integrals at edges away from $\Gamma$ are unchanged, the admissibility 
conditions there are equivalent to those in the vacuum. Conjuncting the appropriate 
admissibility condition at each edge of the lattice we have the following definition 
for spin foam $\Gamma$-admissibility:

\begin{defn}[Spin foam $\Gamma$-admissibility]\label{def:gamma_admis}
Given a Wilson loop $\Gamma$, a spin foam is $\Gamma$-admissible if and only $\forall e \in E$,
edges belonging to $\Gamma$ are $\Gamma$-admissible according to the Definition~\ref{def:edge_charge}, and 
edges not belonging to $\Gamma$ are vacuum admissible according to Definition~\ref{def:edge_vacuum}.
\end{defn}

The above gives a local definition of spin foams contributing to the expectation value of
a Wilson loop observable. The topological structure of the allowed configurations can be
derived by considering (starting from the spin foam with all trivial labels),
first all the ways in which $\Gamma$-admissibility can be satisfied locally at $\Gamma$, and 
then considering the remaining choices\footnote{
As admissibility constraints are additive, this order could be reversed without effecting the
conclusion.}.
To arrive at a $\Gamma$-admissible spin foam, we first satisfy admissibility for each edge on $\Gamma$ by 
specifying a matching there. Depending on how the charged point is matched, this will in turn determine a set of 
plaquettes connected to $\Gamma$. Denote this set of plaquettes by $P^{(0)}$. For each plaquette in $P^{(0)}$,
for each of its edges not attached to $\Gamma$ there is at least one point that has to be routed, as the
spin on the $P^{(0)}$ plaquettes is greater than or equal to $j=\frac{1}{2}$. The situation for a given
edge at this stage of construction is illustrated in Figure~\ref{fig:edgecharge} below. Upon choosing a routing for each 
edge, there is another set $P^{(1)}$ of plaquettes on the other end of the routing. Repeating the above
construction will lead to a set of plaquettes that grows until all routings close, in which
case $P^{(0)} \cup P^{(1)} \cdots$ will define a closed surface of (at least) half-integer charge. 
Note that the growing surface can fold back (not immediately at an edge, but after 
first passing through an edge) on itself; in this case the spins required to define the surface may exceed $j=\frac{1}{2}$
and the surface is said to be self-intersecting\footnote{Or more 
precisely, self-intersecting at a plaquette rather than a vertex or edge; this is the most
useful definition for what will follow.}.

We note that during any step of this construction (and after the surface is closed), an arbitrary labelling
of plaquettes that by themselves satisfy vacuum admissibility conditions at every edge can be added (addition of 
irrep labels at plaquettes) to whatever spins are present by the procedure given above. Moreover, since the
plaquette labelling is edge admissible everywhere, there is at least one and in general multiple edge
labellings that constitute admissible spin foams. Thus, the most general 
solution to the admissibility constraints for a Wilson loop insertion has the form of a direct sum of
a spin foam in the form of a closed (possibly self-intersecting) worldsheet of charge bounding $\Gamma$, and 
a ``background'' spin foam satisfying vacuum boundary conditions.
\begin{figure}
\includegraphics[scale=0.9]{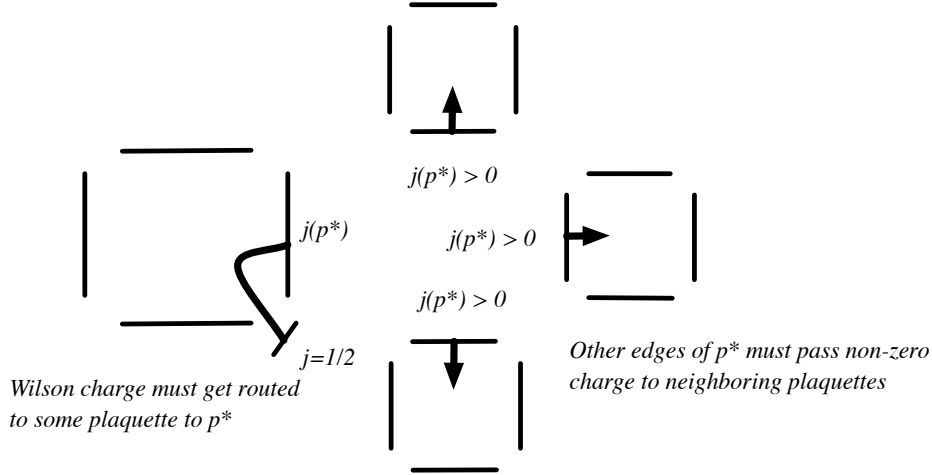}
\caption{Construction and matching conditions for $P^{(0)}$ (three dimensional example)}
\label{fig:edgecharge}
\end{figure}

We hasten to add that while a specific choice of steps in the above construction gives a unique spin foam, 
the converse does not hold. That is, there does not exist a bijection
$\rho : F_{\Gamma} \leftrightarrow F_{V} \times S_{\Gamma}$ where $S_{\Gamma}$ is
the set of closed surfaces connected to $\Gamma$ and $F_{V}$ is the set of vacuum admissible spin foams defined above.
In practical terms, this means a double sum over $\Gamma$-connected worldsheets and over background spin foams would result 
in an overcounting as the same spin foam in $F_{\Gamma}$ would appear multiple times. Further discussion of this
degeneracy is provided in Appendix~\ref{append:degeneracy}.

Notwithstanding, there are several ways to proceed. First, one can simply choose a specific decomposition
$\rho : F_{\Gamma} \rightarrow  F_{V} \times S_{\Gamma}$, and for each worldsheet 
in $S_{\Gamma}$ sum over some restricted set of $F_{V}$ (or vice versa). Unfortunately, it may be difficult to systematically organize such a sum.
However, some work that is particularly relevant to this approach is described by Conrady and Khavkine in~\cite{ConradyKhavkine},
where each spin foam is associated with a unique configuration of fundamental worldsheets that are self-avoiding on a thickened
two-complex associated with the original lattice. This approach appears natural topologically and would be interesting to
develop further.

Another alternative that would allow a direct sum over $\Gamma$-connected worldsheets and vacuum spin foams is
to divide each term by a degeneracy factor equal to the total number of distinct ways the associated spin foam 
can be realized by $F_{V} \times S_{\Gamma}$.

Finally, a (possibly crude) over-estimate for the Wilson loop expectation value could be formed by ignoring
degeneracy when summing over worldsheets and background spin foams.  Because of the possibility of negative
amplitudes in the spin foam model, an absolute value would have to be taken to get a meaningful bound. 
Unfortunately, this over-counting of states combined with taking the absolute value (eliminating
any interference effects) may result in too crude an upper bound to provide a useful confinement result.

The above derivation and considerations are combined in the following statement:
\begin{thm}
Let a function $\rho : F_{\Gamma} \rightarrow F_{V} \times S_{\Gamma}$ be given. Let $S_{\Gamma}^{\rho}(f)$ be the set of
$s \in S_{\Gamma}$ such that $(f,s)$ is in the image of the map $\rho$.  Let $D^{-1}(f,s)$ be the degeneracy
factor associated to the worldsheet-background pair $(f,s)$, as defined in Appendix~\ref{append:degeneracy}. 
Let $S_{\Gamma}(A)$ be the set of worldsheets
with area equal to $A$ in lattice units. Then the following relations hold:
\begin{eqnarray}
\left< O_{\Gamma_{w}} \right> &=& \frac{\sum_{f \in F_{\Gamma}} A_{\Gamma}(f)} {\sum_{f \in F_{V}} A(f)} \\ \nonumber
 &=& \frac{\sum_{f \in F_{V}} \left(\sum_{s \in S_{\Gamma}^{\rho}(f)} I(f,s) \right) A(f) }{\sum_{f \in F_{V}} A(f)} \\ \nonumber
  &=& \frac{\sum_{f \in F_{V}} \left(\sum_{s \in S_{\Gamma}} D^{-1}(f,s)I(f,s) \right) A(f) }{\sum_{f \in F_{V}} A(f)} \\ \nonumber
  &=& \frac{\sum_{f \in F_{V}} \left[ \sum_{A \geq A_{\text{\emph{Min}}}(\Gamma)} \sum_{s \in S_{\Gamma}(A)} D^{-1}(f,s)I(f,s) \right] A(f) }{\sum_{f \in F_{V}} A(f)} 
\end{eqnarray}
where
\begin{equation}\label{interact}
I(f,s)\equiv
\prod_{p \in s}\frac{A_{P}(f(p),s(p))}{A_{P}(p)}
\prod_{e \in s}\frac{\bar{A}_{E}(f(e),s(e))}{A_{E}(f(e))}
\prod_{v \in s}\frac{\bar{A}_{V}(f(v),s(v))}{A_{V}(f(v))},
\end{equation}
is the \emph{interaction factor} that measures the change in amplitude associated with inserting a 
worldsheet $s$ into a background spin foam $f$ to obtain a new spin foam $f'$. The numerators
of~(\ref{interact}) are evaluated on the spin foam $f'=f+s$, and can be thought of as 
functions of the original spin foam variables displaced by the presence of the worldsheet $s$.
The functions $f(p)$, $f(e)$ and $f(v)$ pick out the labels assigned by the spin foam to a plaquette,
edge or vertex, respectively.
\label{decompthm}
\end{thm}

The first equality is simply the spin foam expression for Wilson loop expectation values as a ratio
of partition functions built from $\Gamma$-admissible and vacuum spin foams.

In the second equality, a sum is performed over all background spin foams. For each
background spin foam, a subset $S_{\Gamma}^{\rho}(f)$ of all $\Gamma$-attached worldsheets are summed
over; in general this is a proper subset of $S_{\Gamma}$ to avoid overcounting.
In the third equality, an \emph{unrestricted} double sum over all background spin foams and all worldsheets is performed,
with an inverse degeneracy factor $D^{-1}(f,s)$ included to correct the overcounting.
In the final line, we have organized the sum over $\Gamma$-terminated worldsheets by their surface
area (in plaquette units). The sum starts at $A_{\text{Min}}$, the area
of the smallest closed surface that terminates on $\Gamma$.

We should emphasize several interesting properties of the interaction factor $I(f,s)$ that appears above.
First, we observe that (for a given worldsheet $s$) it is an observable on the vacuum ensemble. Although
division occurs in its definition, on the space of admissible vacuum spin foams the observable is always finite.
Moreover, as is manifest from its definition it factors into a product of local terms, and in particular
the plaquette term can be readily found in closed form.
With the above in hand, we now turn to a possible confinement scenario.

\section{Confinement at Weak Coupling: A Scenario}
We describe here a general scenario for confinement using the above reformulation of the Wilson loop criteria
in terms of worldsheets interacting with background spin foams from the vacuum ensemble. To begin, we 
introduce the notion of confinement relative to a fixed ``background'' spin foam from $F_{V}$, the set
of vacuum spin foams. A first important step is to distinguish the plaquette and vertex parts of the interaction factor:
\begin{eqnarray}
I_{V}(f,s) \equiv \prod_{v \in V} \frac{ \bar{A}_{V}(f(v),s(v)) } { A_{V}(f(v)) }, \qquad I_{P}(f(p),s(p)) \equiv \prod_{p \in P} 
\frac{ \bar{A}_{P}(f(p),s(p))}{A_{P}(f(p))}.
\end{eqnarray}
We assume the edge amplitude can be absorbed into the vertex amplitude, which can be generically achieved by
including a square root of the edge amplitude on either vertex of a given edge; with this convention the 
interaction term factors as $I(f,s)=I_{V}(f,s)I_{P}(f,s)$. From Theorem~\ref{decompthm}, we have
\begin{eqnarray}\label{area_sumform}
 \left< O_{\Gamma_{w}} \right>  
 &=& \frac{\sum_{f \in F_{V}} \left[ \sum_{A \geq A_{\text{Min}}(\Gamma)} \sum_{s \in S_{\Gamma}(A)} D^{-1}(f,s)I(f,s) \right] A(f) }{\sum_{f \in F_{V}} A(f)},
\end{eqnarray}
where we have used the resummation by area that will be particularly well-suited for comparing effective tensions in
what follows.

We shall next present a sequence of arguments for how the above expression may lead to confining behavior in the 
weak coupling ($\beta \rightarrow \infty$) limit.  Our first step is to provide for the existence of
$\left< O_{\Gamma_{w}} \right>$ at a fixed $\Gamma$ by identifying the conditions under which it
is bounded from above by a convergent exponential series in the worldsheet area $A$. We then compare the leading order 
terms for $\Gamma$ of different size (minimal area) and propose a condition under which 
they are bounded by an exponential decay in minimal area.

\subsection{Area damping from plaquette interaction} An elementary analysis of the plaquette interaction
factor $I_{P}$ for the $SU(2)$ case (see~\cite{Conf2} for details --- note we are using whole rather
than half integers to label irreps) with the heat kernel action~\cite{Menotti} provides a plaquette
interaction that has the form 
\begin{eqnarray}\label{plaq_tension}
I_{P}(f,s) &=& \left(e^{\beta^{-1} \sum_{p \in s} \delta_{p}(s) j_{p} } \right)
\left( e^{-(2\beta)^{-1}\sum_{p \in s} \delta_{p}(s)(\delta_{p}(s)+2)} \right)
\left(\prod_{p \in s} \frac{j_{p}+1+\delta_{p}(s)}{j_{p}+1} \right)  \\ \nonumber
           &\equiv& e^{-\tau_{p}(f,s)A(s)},
\end{eqnarray}
where the sum is over every plaquette in the worldsheet $s$ and we define the
\emph{plaquette tension} as $\tau_{p}(f,s) \equiv
\frac{ \log \left( I_{P}(f,s) \right)}{A(s)}$ where $A(s)$ is the lattice area of $s$
and $\delta_{p}(s)$ is the irrep associated to the plaquette
by the worldsheet. Note the second bracket factor in~(\ref{plaq_tension}) goes to unity 
as $\beta \rightarrow \infty$; consequently, this factor does not play the crucial
role that it does in the \emph{strong} coupling limit. Given that the third factor
can be made arbitrarily large on any background on surfaces of high enough self-intersection,
how can a positive tension be provided?
We see by inspection that if the $j_{p}$ (on the backgrounds that dominate $Z_{V}$ at
any given $\beta$) grows at the same rate or faster than $\beta$ as $\beta \rightarrow \infty$, 
than the first factor will be less than unity (provide tension to each worldsheet) on
those backgrounds. Moreover, it will also damp the third factor (exponential decay in $\delta$ vs
linear growth in $\delta$) for sheets of arbitrarily high self-intersection.
To continue our analysis to the rest of the interaction factor we substitute
our exponentiated form of the plaquette interaction~(\ref{plaq_tension}) into
the expectation loop expression~(\ref{area_sumform}) above:
\begin{eqnarray}\label{plaq_subs}
\left< O_{\Gamma_{w}} \right> 
= \frac{\sum_{f \in F_{V}} \left[ \sum_{A \geq A_{\text{Min}}(\Gamma)} \sum_{s \in S_{\Gamma}(A)} D^{-1}(f,s)e^{-\tau_{p}(f,s)A}
I_{V}(f,s) \right] A(f) }
{\sum_{f \in F_{V}} A(f)}.
\end{eqnarray}
For a given $f$, an infinite number of above unity interaction terms (the summands appearing in square brackets) 
in~(\ref{plaq_subs}) above would ruin convergence, or at best give conditional convergence and 
associated ambiguities. Thus, we would like to have below unity interaction factors on all but a 
finite set of sheets (at least on those background $f$ that aren't suppressed in the weak coupling
limit).

Inspection of~(\ref{plaq_subs}) reveals further conditions: $\tau_{p}(f,s)$ at large $\beta$ must not only be 
positive, it must be sufficient to overcome the growth in the number
of surfaces of a given area, which also grows exponentially (is bounded from above) in the area~\cite{IDV1} with some associated
negative tension\footnote{Although we suppress it in the notation $T_{0}=T_{0}(A)$ carries some $A$ dependence.}
, i.e. as $e^{T_{0}(A-A_{\text{Min}})}$. We note in passing that $T_{0}$ is dimension dependent and so in principle
may lead to different tensions or even departures from confining behavior depending on dimension.

As the degeneracy factor is always less than unity, the remaining tension after accounting for the growth in surface number $e^{T_{0}A}$ in each area class has to be sufficient to overcome any above unity interaction factors from 
the vertex amplitude.  To isolate the tension coming from the plaquette amplitude, we consider a related model where
$I_{V} \equiv 1$ (note that $A_{V}$ in $A(f)$ is not set to unity) and make the following definition:
\begin{defn}[Area damping for plaquette interaction]
A plaquette interaction (and the plaquette amplitude from which it is derived) is said to be \emph{area damping} if
\begin{equation}
\frac{\sum_{f \in F_{V}} \left[ \sum_{A \geq A_{\text{Min}}(\Gamma)} \sum_{s \in S_{\Gamma}(A)} D^{-1}(f,s)e^{-\tau_{p}(f,s)A}
\right] A(f) }
{\sum_{f \in F_{V}} A(f)} < \infty.
\end{equation}
\end{defn}
As both vertex and plaquette interactions factor over the area of a worldsheet, what is gained by 
analyzing them separately? In the $G=SU(2)$, $D=3$ case it was found in~\cite{Conf2} that $I_{V}(f,s) \rightarrow 1$ for 
large spin backgrounds (characteristic of the weak coupling limit) and self-avoiding worldsheets.
Consequently $I_{P}$ rather than $I_{V}$ would have to provide confinement for this class of worldsheets and
backgrounds. The large $\beta$ form of $I_{P}$~(\ref{plaq_tension}) provides such a means to provide effective 
tension, assuming the typical spin grows fast enough with increasing $\beta$ as mentioned above.  
This motivates our conjecture that for Yang-Mills theories in general gauge groups and dimensions,
it is the $I_{P}$ that provides an effective tension as described above. If this conjecture is true,
than area damping is a necessary condition for the existence of the expectation value.
Given the area damping condition, the behavior of the vertex interaction becomes decisive; we turn
to this next.

\subsection{Stability of vertex interaction}
For the area damping from plaquette interaction to provide confinement within our proposed scenario,
the behavior of the vertex amplitude that would lead to negative overall tension from (above unity interaction factors)
must be bounded from above to a degree determined by the other tensions (on all but a finite set of sheets).
We start by making the following definition:
\begin{eqnarray}
 I_{V}(f,s) \equiv \epsilon(f,s) \mathcal{P}(f,s) e^{-\tau_{v}(f,s)A(s)},
\end{eqnarray}
where $\epsilon(f,s)$ is the overall sign of vertex interaction factor, the quantity $\mathcal{P}(f,s)$ is
the \emph{perimeter factor}\footnote{
We factor out a perimeter factor as the vertex amplitudes have a different form along $\Gamma$,
as discussed above; this is often of practical utility when working with a specific model.
}
defined as $\mathcal{P}(f,s) = |\prod_{v \in \Gamma} I_{V}(f,s)|$ and $\tau_{v}$ is
defined implicitly by the above. Unlike the plaquette factor, we note $I_{V}(f,s)$ can in principle introduce a
negative sign, as it involves spin network evaluations that are not strictly positive. 

We next introduce effective area-averaged tension $\bar{\tau}(f,s)$ as
\begin{equation}
e^{-\bar{\tau}(f,A)A} \equiv \frac{\sum_{s \in S_{\Gamma}(A)}I_{V}(f,s)I_{P}(f,s)}{e^{T_{0}(A-A_{\text{Min}}) }} =
\frac{\sum_{s \in S_{\Gamma}(A)}D^{-1}(f,s)\epsilon(f,s)\mathcal{P}(f,s)
e^{-(\tau_{p}(f,s)+\tau_{v}(f,s))A}}{e^{T_{0}(A-A_{\text{Min}})}}.
\end{equation}
To compare the this effective tension from the interaction factor with the ``negative'' tension
arising from growth in surface number, we consider the following form of the expectation value:
\begin{eqnarray}\label{ansatz}
\left< O_{\Gamma_{w}} \right> =
\frac{\sum_{f \in F_{V}} \left[ e^{-T_{0}A_{\text{Min}}} \sum_{A \geq A_{\text{Min}}}
e^{ \left( T_{0}-\bar{\tau}(f,A) \right) A} \right] A(f)}
{\sum_{f \in F_{V}}A(f)},
\end{eqnarray}
which suggests the inequality
\begin{eqnarray}\label{inequality}
\bar{\tau}(f,A)-T_{0} > 0
\end{eqnarray}
as a necessary condition for a contribution to the sum over $A$ to be below unity (for a given background spin foam $f$). 
If this inequality is satisfied for all $A$ for a foam $f$ (and the infinum of $\bar{\tau}$
over all the $A$ is non-zero), then the vacuum amplitude $A(f)$ is multiplied by a sum of
terms having the form of a constant of order $e^{-T_{0}A_{\text{Min}}}$, and a factor exponentially
decaying at a rate proportional to the area. This leads to the following definition:
\begin{defn}[Vertex Stability]
The set of ordered pairs of $(f,A)$ where inequality~(\ref{inequality}) holds is said to exhibit \emph{vertex stability}.
\end{defn}
It may be the case that the inequality is violated on some finite subset of areas --- in which
case the sum of interaction factors still converges. Additionally, if there are any $f \in F_{V}$ for 
which there are areas for which inequality violation occurs, they should make vanishing contribution 
to the expectation value in the large $\beta$ limit. If this were not the case it may be very difficult 
(although not impossible) to organize the leading order of the sums to find an area law. An example of 
suppression of such ill-behaved world-sheet background sectors can in fact be shown for the ``sub-asymptotic''
configurations (essentially those containing low spin) in the $D=3, G=SU(2)$ model~\cite{Conf2}.

\subsection{Balance conditions and a scenario for confinement}
The weak coupling limit $\left< O_{\Gamma_{w}} \right>$ presented in~(\ref{ansatz}) subject to
the inequality~(\ref{inequality}) is highly suggestive of a convergent interaction factor
and a possible area law. That is, if $A(f)$ were strictly non-negative one could form an 
upper bound with the form of a series of decaying exponentials in the area, and compare
the leading order contribution associated to $\Gamma$ of increasing size.
Unfortunately, positivity does not hold for $A(f)$ in a general spin foam model for Yang-Mills.
Hence, before proceeding further we must deal with the non-positivity of the amplitude.
One approach to account for this is to write the ansatz~(\ref{ansatz}) above as follows:
\begin{eqnarray}\label{ansatz_balanced}
\left< O_{\Gamma_{w}} \right> &\rightarrow& 
\frac{\sum_{f \in F_{V}} \left[e^{-T_{0}A_{\text{Min}}} \sum_{A \geq A_{\text{Min}}}
e^{ \left( T_{0}-\bar{\tau}(f,A) \right) A} \right] A(f)}
{\sum_{f \in F_{V}}A(f)} \\ \nonumber
&=& 
\frac{ \left( \sum_{f \in F_{V}^{+}} - \sum_{f \in F_{V}^{-}}\right) \left[e^{-T_{0}A_{\text{Min}}}
\sum_{A \geq A_{\text{Min}}}
e^{ \left( T_{0}-\bar{\tau}(f,A) \right) A}\right] |A(f)|}
{\sum_{f \in F_{V}}A(f)} \\ \nonumber
&=&
\left( \sum_{f \in F_{V}^{+}}|A(f)| \right) \left( \frac{\sum_{f \in F_{V}^{+}} 
e^{-T_{0}A_{\text{Min}}}\sum_{A \geq A_{\text{Min}}}
e^{ \left( T_{0}-\bar{\tau}(f,A) \right) A}
|A(f)|}{\sum_{f \in F_{V}^{+}}|A(f)|} \right) Z_{V}^{-1}\\ \nonumber
&-& 
\left( \sum_{f \in F_{V}^{-}}|A(f)| \right) \left( \frac{\sum_{f \in F_{V}^{-}} 
e^{-T_{0}A_{\text{Min}}} \sum_{A \geq A_{\text{Min}}}
e^{ \left( T_{0}-\bar{\tau}(f,A) \right) A}
|A(f)|}{\sum_{f \in F_{V}^{-}}|A(f)|} \right) Z_{V}^{-1}
\end{eqnarray}
where we have used the fact that $\sum_{f \in F_{V}}A(f)$ and $\sum_{f \in F_{V}}A_{\Gamma}(f)$ 
are absolutely convergent\footnote{
This can be shown for any given $\beta$ due to upper bounds on the vacuum and $\Gamma$ vertex
amplitudes.
}
and can thus be broken into two sums
$\sum_{f \in F_{V}}A(f)= \sum_{f \in F_{V}^{+}}|A(f)| - \sum_{f \in F_{V}^{-}}|A(f)|$.

We next make the following conjecture: \emph{the interaction factor for any given worldsheet $s$, averaged over negative
amplitude states $F_{V}^{-}$ approaches its average over positive amplitude states $F_{V}^{+}$ as $\beta \rightarrow \infty$}.
That is, 
\begin{eqnarray}
\left< \Sigma \right>_{+} &\equiv&
\left( \frac{\sum_{f \in F_{V}^{+}} 
e^{-T_{0}A_{\text{Min}}} \sum_{A \geq A_{\text{Min}}}e^{ \left( T_{0}-\bar{\tau}(f,A) \right) A}
|A(f)|}{\sum_{f \in F_{V}^{+}}|A(f)|} \right) \\ \nonumber
&\rightarrow& \left( \frac{\sum_{f \in F_{V}^{-}} 
e^{-T_{0}A_{\text{Min}}} \sum_{A \geq A_{\text{Min}}}
e^{ \left( T_{0}-\bar{\tau}(f,A) \right) A}
|A(f)|}{\sum_{f \in F_{V}^{-}}|A(f)|} \right) \quad \text{as $\beta \rightarrow \infty$.} \\ \nonumber
&\equiv& \left< \Sigma \right>_{-}
\end{eqnarray}
As an immediate consequence we then have from~(\ref{ansatz})
\begin{eqnarray}
\left< O_{\Gamma_{w}} \right> \rightarrow
\left< \Sigma \right>_{+} 
\left(
\frac{  \sum_{f \in F_{V}^{+}} |A(f) - \sum_{f \in F_{V}^{-}} |A(f)|}
     {  \sum_{f \in F_{V}^{+}} |A(f) - \sum_{f \in F_{V}^{-}} |A(f)|} 
\right)
= \left< \Sigma \right>_{+}
\end{eqnarray}
which allows us to write
\begin{eqnarray}\label{final_bound}
\left< O_{\Gamma_{w}} \right>
&\rightarrow&
\frac{\sum_{f \in F_{V}^{+}}
e^{-T_{0}A_{\text{Min}}}
\sum_{A \geq A_{\text{Min}}}
e^{ \left( T_{0}-\bar{\tau}(f,A) \right) A}
|A(f)|}{\sum_{f \in F_{V}^{+}}|A(f)|}
\\ \nonumber
&<& 
\frac{
\left(
e^{-T_{0}A_{\text{Min}}}
\sum_{A \geq A_{\text{Min}}}e^{-\tau^{*}A}
\right)
\sum_{f \in F_{V}^{+,\tau}} |A(f)|} {\sum_{f \in F_{V}^{\tau^{*}+}}|A(f)|} + E(\beta) \\ \nonumber
&=& e^{-T_{0}A_{\text{Min}}} \sum_{A \geq A_{\text{Min}}}e^{-\tau^{*}A} + E(\beta)
\\ \nonumber
\end{eqnarray}
where $\tau^{*} = \text{inf}_{F_{V}^{\tau^{*}+},A} \left[ \bar{\tau}(f,A)-T_{0} \right]$
is the infinum (greatest lower bound) of the effective tension over areas and $F_{V}^{\tau^{*}+}$,
the set of positive amplitude vacuum spin that give positive tension; it is assumed to exist and is non-zero\footnote{
Otherwise terms of arbitrarily small tension would be present in the continuum limit; the theory
would not be confining.}. The term $E(\beta)$ represents contributions which may have zero or negative tension but
go to zero as $\beta \rightarrow \infty$.  An example of such are sub-asymptotic
contributions discussed in~\cite{Conf2}. In a theory where confinement is true, we refer to $F_{V}^{\tau^{*}+}$
as the $\emph{dominant sector}$.

A final step in showing confinement is to compare the leading order terms in $\left< O_{\Gamma_{w}} \right>$
for different loops $\Gamma$ of increasing size (lattice area of minimal surface). 
A useful construct is to consider the series of background averaged expectation values:
\begin{eqnarray}
\alpha_{\Gamma}(A) = \frac{\sum_{f \in F_{V}^{+}}
e^{ \left( T_{0}-\bar{\tau}(f,A) \right) A}
|A(f)|}{\sum_{f \in F_{V}^{+}}|A(f)|},
\end{eqnarray}
in terms of which we can write
\begin{eqnarray}
\left< O_{\Gamma_{w}} \right> = e^{-T_{0}A_{\text{Min}}} \sum_{A \geq A_{\text{Min}}} \alpha_{\Gamma}(A).
\end{eqnarray}
We next need a way of comparing $\alpha_{\Gamma}(A_{\text{Min}})$ for $\Gamma$ of increasing size. 
Suppose we make a generalization of the infinum $\tau^{*}$ used in~(\ref{final_bound}). As $\tau^{*}$ in that
case is defined for a specific $\Gamma$, we write $\tau^{*}(\Gamma)$ and take the infinum over some set
of loops $\Gamma$ of interest, $\mathcal{G}$:
\begin{equation}
\hat{\tau}_{\mathcal{G}} \equiv \text{inf}_{\mathcal{G}}[\tau^{*}(\Gamma)].
\end{equation}
If $\hat{\tau} > 0$, then the following relation will be true:
\begin{equation}\label{compare_loops}
\alpha_{\Gamma_{1}}(A_{\text{Min}}(\Gamma)) < e^{-\hat{\tau}_{\mathcal{G}}A_{\text{Min}}} \quad
\text{for all $\Gamma \in \mathcal{G}$}.
\end{equation}
Note that the condition~(\ref{compare_loops}) by itself permits all the manner of fluctuations in
$\left< O_{\Gamma_{w}} \right>$ as $\Gamma$ is increased, as long as they all take place within a
decaying envelope with fall-off given by $\hat{\tau}$. It is however not difficult
to show monotonic decrease in the $\alpha_{\Gamma}(A)$, both for a fixed $\Gamma$ and varying $A$ and
for varying $\Gamma$. Such a result puts additional bounds on the extent of fluctuations in the $\bar{\tau}$.

\subsection{Critical Remarks}
The area damping and vertex stability conditions represent non-trivial statements regarding
the spin foam form of the gauge theory in the weak coupling limit.  For any given model,
further analysis will be required to examine the validity of these propositions.
Notwithstanding, if these proposals prove false the alternatives for realizing confinement
are quite constrained. We briefly mention a few of the possibilities here.

It may be the case that for certain gauge groups, the area damping condition fails, i.e. 
the plaquette amplitude provides insufficient tension relative to the surface
growth factor on backgrounds that dominate the $\beta \rightarrow \infty$ limit. 
If this ``vertex damping'' scenario is true, it would be quite remarkable as it would 
place very specific constraints on displacements of vertex amplitudes with arbitrarily 
large spin arguments, whose contribution to $Z_{V}$ grows as $\beta \rightarrow \infty$.  

A second scenario more closely related to the one proposed is that the plaquette and vertex amplitudes are
by themselves insufficient but combined can overcome the surface number growth.  Because neither
type of interaction factor is alone sufficient, one could term this a ``cooperative'' confinement scenario.

In considering the final steps of the above, the balance condition calls for particular
scrutiny, as it was motivated largely for its convenient mathematical properties. 
If it proves to be false, a weakened form may still be true. In this weakened form,
the balance condition may hold true only for a subset of spin foams, in which case the 
remaining spin foams would have to be shown to be suppressed or exhibit area law behavior
by some other line of argument.
If such balance conditions prove false (or impractical to show), it may be fruitful to
consider other means of passing from the form~(\ref{ansatz}) to an area law
inequality, such as by employing a stationary phase analysis. Such an analysis may be
particularly appropriate in the asymptotic setting where analytic formulas for vertex
amplitudes are available, as in~\cite{Conf2}.

The presence of the degeneracy factor in defining the effective tension is also 
somewhat unsatisfying, as we at present don't have
a good estimate for it. Thus, we don't \emph{a priori} know that configurations with 
large vertex interactions which might otherwise spoil confinement aren't damped
by the degeneracy factor. A particularly well-chosen decomposition $\rho$ (see 
Theorem 3.1 for definition) might offer a better understanding and is the topic of ongoing work by the author.
Notwithstanding, if sufficiently strong bounds on the behavior of vertex interactions
can be found, they may be sufficient (combined with estimates on other interaction factors)
to provide at least an area law bound. Such a bound likely would not estimate
the actual tension very well, due to a lack of degeneracy factor estimate 
(other than that it is less than unity).

A final comment relates to our treatment of the perimeter factors.
Our analysis does not rule out a non-trivial perimeter law, as meeting the conditions
of area damping, vertex stability, and the existence of the infinum over $\Gamma$
yields an area-law bound (rather than equality).

\section{Conclusions}
Based on some elementary considerations that allow Wilson loop expectation values to be expressed as a double
sum over worldsheets and vacuum spin foam backgrounds (modulo a degeneracy factor), we have attempted to
map out some of the properties we expect to hold at weak coupling in order to provide an area-law decay.
Specifically, the concepts introduced in this paper can summarized as follows:
\begin{enumerate}
\item{Wilson loop expectation values can be expressed as a certain type of observable having the form of
a sum over worldsheets bounded by $\Gamma$. Each worldsheet contributes an interaction factor that
is a function of variables local to the worldsheet.}
\item{The interaction factors into a plaquette interaction and a vertex interaction. It is conjectured that
the large $\beta$ form of the plaquette interaction provides the critical amount of tension required
for area law behavior.}
\item{
An inequality relating the effective tension provided by interaction factor and the worldsheet number 
in each area class is conjectured to hold on vacuum spin foams that dominate the large $\beta$
limit. The contribution to the expectation value of those spin foams and area classes 
where the inequality is violated must be shown to go to zero as $\beta \rightarrow \infty$.
}
\end{enumerate}
Put in more descriptive terms, we propose that a necessary feature for confinement at weak coupling 
is that the typical vacuum configurations at large $\beta$ exhibit sufficiently high spin relative
to $\beta$ such that the interaction factor (a product of both plaquette and vertex interaction factors)
averaged over worldsheets of a given area provides an effective tension greater than the ``negative tension''
provided by the ``entropy'' (number of) lattice worldsheets bounding $\Gamma$ of that area.

Qualitatively, this mechanism is suggestive of the dual superconductor model in which
the Yang-Mills vacuum ``expels'' electric flux.  It may be possible to develop this connection further ---
spin foams can be thought of as path integrals over spin networks, which provide a complete basis
for the physical Hilbert space of the lattice regularized gauge theory~\cite{BDMUV,BDMUV2}.  Intriguingly,
in the spin network basis, the ``electric'' part of the lattice Hamiltonian is diagonal~\cite{BDMUV,BDMUV2} and 
the eigenvalues are products of Casimirs in the irrep labels, which appear explicitly in the heat kernel
form of the plaquette amplitude which leads to the plaquette interaction factor. We expect that, 
roughly speaking, the plaquette amplitude arises from the electric part of the lattice action, however 
further work on the relation between Hamiltonian and Lagrangian formalisms is likely required to make 
this statement more precise.

At this point we would like to emphasize some differences with the confinement at strong coupling. In going from 
strong to weak coupling the confinement mechanism shifts from a high tension worldsheet embedded in sparse
(low-spin) backgrounds, to that of a low tension worldsheet that is forced (again, by virtue of
having to end on $\Gamma$) to interact with vacuum spin foam where high spins form a percolating
background of spin which grows with increasing $\beta$.  In the weak coupling limit, partially sub-asymptotic configurations 
(that can exhibit negative tension through the vertex interactions) become exponentially suppressed,
and thus don't spoil the confinement.

Due to the limit nature of these arguments, which take place in the opposite extremes of strong 
and weak coupling, an intermediate region of ``roughening'' is not ruled out in regions
of the phase diagram (e.g. coupling and lattice size) in which neither strong or weak mechanisms are 
sufficient to provide a substantial tension. Through the use of dual spin foam algorithms, it should be possible
to quantify the relative influence of these two mechanisms at a given coupling.

To further assess the validity of this proposed confinement mechanism, a number of additional 
results will be needed, the details of which depend on the space-time dimension and gauge group
considered. While the case of $SU(2)$ in three space-time dimensions taken up in~\cite{Conf2} seems 
so far consistent with the proposed picture, the case of $SU(2)$ on a four-dimensional lattice remains to 
be considered in more detail. A formula for the four-dimensional vertex amplitude has recently been 
given in~\cite{CC2009}; being built up of a sum of $6j$ and $3j$ symbols it may be 
possible to find a suitable asymptotic description simply by substitution, or else through a
more direct approach. In both three and four dimensions, more careful
control of the asymptotic expression will be required to handle the case of self-intersecting worldsheets 
interacting with asymptotic spin foams.  

To extend the analysis to $SU(3)$, suitable asymptotic formulas for this case will be required
in order to show that the vertex interaction trivializes or at least provides some bounded amount of 
negative tension. To the author's knowledge, asymptotic formulae analogous to Ponzano-Regge asymptotics
for $SU(2)$ $6j$ symbols have not yet been found.  In general, the computation of $6j$ and $3j$ symbols and 
their recoupling theory is considerably less developed than the $SU(2)$ case at this time.  However, we 
hope that some of the work to date~\cite{A1992,KIM2006,K97} will provide a strong starting point for 
eventually understanding the asymptotics of $SU(3)$ spin networks.

While the above problems may present considerable technical challenges, we believe that the main
result of our work can be stated clearly: assuming that Yang-Mills theory truly describes our world and that
confinement is true, the weak coupling behavior of vertex interactions and plaquette interactions
relative to the growth rate in surface number with area are subject to non-trivial constraints.
These constraints suggest very specific lines of inquiry, some of which are identified in here and in~\cite{Conf2}.

In pursuing a deeper understanding of these constraints, we are encouraged by some powerful new
methods and results involving the asymptotics of $6j$ symbols and spin networks, from both the mathematics
literature~\cite{GVdW09,AA09,LJ09} and in the context of the spin foam program for 
quantum gravity~\cite{FNR06,GU08,LD09}, for which the asymptotics of spin foam amplitudes 
also play a crucial role.

\begin{acknowledgement*}
The author would like to thank Dan Christensen, Florian Conrady, and Igor Khavkine for valuable discussions that
have influenced this work.
\end{acknowledgement*}

\begin{appendices}

\section{Degeneracy of Spin Foam Decompositions}\label{append:degeneracy}
\begin{figure}
\includegraphics[scale=0.27]{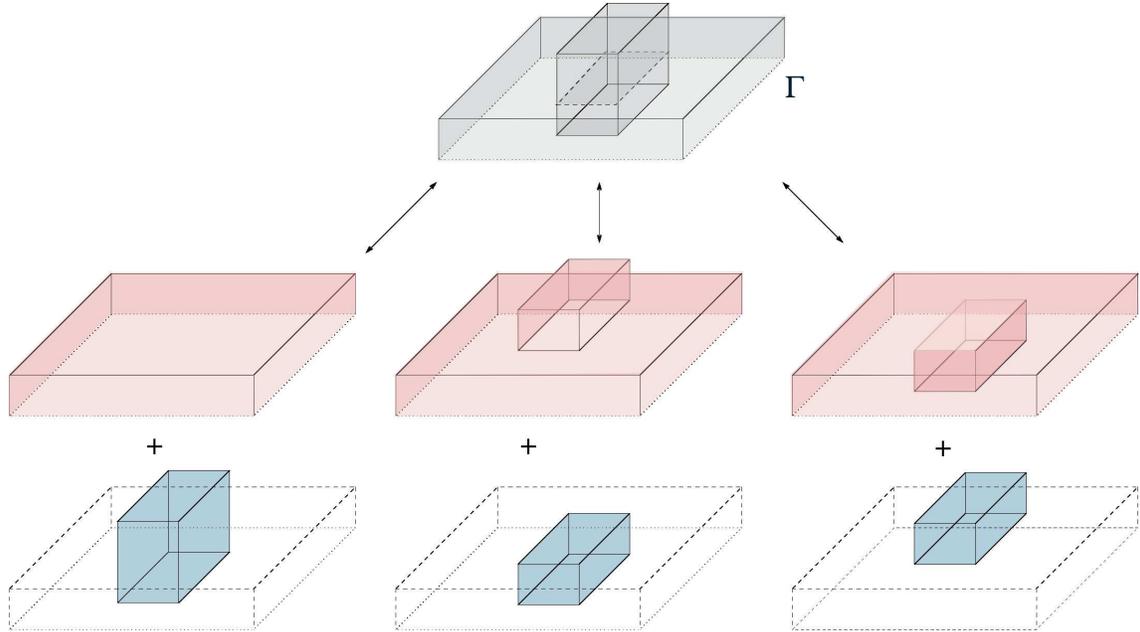}
\caption{Multiple decompositions of the same $\Gamma$-admissible spin foam (top) into worldsheet (middle) plus background
(bottom).}\label{fig:degen_basic}
\end{figure}
As discussed above, it is particularly natural to view contributions to $\left< O_{\Gamma_{w}} \right>$ as
composed of a worldsheet bounding $\Gamma$ together with some ``background'' spin foam associated
with the undisturbed Yang-Mills vacuum (no external charge).  A summation over all worldsheets
and all backgrounds will (quite seriously in most cases) over-count the underlying configurations $F_{\Gamma}$.

A simple example of this degeneracy is the spin foam shown in Figure~\ref{fig:degen_basic}.
At the top is illustrate a $\Gamma$-admissible spin foam. It can be decomposed as a worldsheet connected
to $\Gamma$ (dotted line) which intersects a closed surface belonging to $F_{V}$. Alternatively,
it can be decomposed as the direct sum of a worldsheet raised upwards in a way that conforms with the boundary
of the box and a smaller closed surface.  Finally, the configuration where the worldsheet is pulled downwards and
a smaller closed surface is fit over top is also equivalent. 

In the above case, a $\Gamma$-admissible spin foam had three separate decompositions. This leads us to the
following definition:
\begin{defn}[Degeneracy factor] Let $\gamma$ be a $\Gamma$-admissible spin foam. As discussed in Section 3, there
exists at least one ordered pair of vacuum spin foams $f$ and $s$ such that $\gamma = f + s$. Define the \emph{degeneracy
factor} to be the number of distinct such decompositions for a given $\gamma$.
\end{defn}
We note in passing the $D(\gamma)$ can also be written as $D(f,s) \equiv D(f+s)$ and because it is non-zero the
function $D^{-1}(f,s)$ is always well-defined.

\begin{figure}
\includegraphics[scale=0.14]{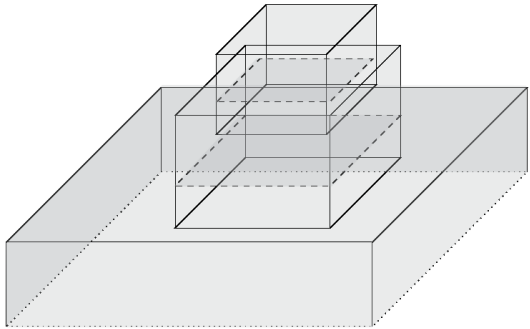}
\caption{Example of a pyramid-like $\Gamma$-admissible spin foam leading to a recursive set of degeneracies.}\label{fig:pyramid}
\end{figure}

In theory, one may attempt to compute a degeneracy factor by finding how many closed background spin foams are sliced
``in half'' by the worldsheet. The pyramid-like construct shown in Figure~\ref{fig:pyramid} illustrates how this 
may be difficult, as  additional degeneracies can be generated from an initial slice.

\section{Mass Gap and Generalized Wilson loop criteria}\label{append:massgap}
We give here the following (standard) definition of the mass gap observable:
\begin{defn}[Mass gap observable]\label{def:mgap_obs}
\begin{equation}
\left< O_{ \Gamma_{1} \Gamma_{2}(x) } \right> \equiv \left< O_{\Gamma_{1}}O_{\Gamma_{2}} \right> - \left< O_{\Gamma_{1}} \right> 
\left< O_{\Gamma_{2}} \right>
\end{equation}
where $\Gamma_{1}$ and $\Gamma_{2}$ are the boundaries of two parallel plaquettes (one above the other) separated by distance
$x$. The one-point expectation values $\left< O_{\Gamma_{1}} \right>$ and $\left< O_{\Gamma_{2}} \right>$
are defined as above with $\Gamma_{w}=\Gamma_{i}$. The product expectation value $\left< O_{\Gamma_{1}}O_{\Gamma_{2}} \right>$
is defined as follows:
\begin{equation}
\left<O_{\Gamma_{1}}O_{\Gamma_{2}} \right>  \equiv \frac{Z_{\Gamma_{1} \Gamma_{2}}}{Z_{V}} \equiv \frac{\int \left( \prod_{e\in E}dg_{e} \right) \,
\text{Tr} \left( \prod_{e \in \Gamma_{1}} D_{w}(g_{e}) \right) \text{Tr} \left( \prod_{e \in \Gamma_{2}}D_{w}(g_{e}) \right) e^{-\sum_{p\in P}S(g_{p})}}
{\int \left( \prod_{e\in E}dg_{e} \right)\, e^{-\sum_{p\in P}S(g_{p})}}.
\end{equation}
\end{defn}
As in the case of confinement (where the Wilson loop consisted of a single connected component), an analysis of admissibility at the
two components $\Gamma_{1}$ and $\Gamma_{2}$ allows contributions $Z_{\Gamma_{1} \Gamma_{2}}$ to be resolved into a background
spin foam upon which are superimposed one or more surfaces bounding the $\Gamma$. Specifically, we can write a sum of 
the form given in Theorem~\ref{decompthm}:
\begin{eqnarray}
\left< O_{\Gamma_{1} \Gamma_{2}(x)} \right> &=& \frac{\sum_{f \in F_{\Gamma_{1} \Gamma_{2}(x)}} A_{\Gamma}(f)} 
  {\sum_{f \in F_{V}} A(f)} \\ \nonumber
  &=& \frac{\sum_{f \in F_{V}} \left(\sum_{s \in S_{\Gamma_{1} \Gamma_{2}(x)}^{\rho}(f)} I(f,s) \right) A(f) }
  {\sum_{f \in F_{V}} A(f)} \\ \nonumber
  &=& \frac{\sum_{f \in F_{V}} \left(\sum_{s \in S_{\Gamma_{1} \Gamma_{2}(x)}} D^{-1}(f,s)I(f,s) \right) A(f) }
  {\sum_{f \in F_{V}} A(f)}, 
\end{eqnarray}
where the only difference with the case of a single connected Wilson loop component is that the set $S_{\Gamma}$ has to be suitably
generalized to account for the disconnected topology of $\Gamma$.  One finds that $S_{\Gamma}$ for a two-point function consists of
the following types:
\begin{enumerate}
\item An (arbitrary) worldsheet with both $\Gamma_{1}$ and $\Gamma_{2}$ as boundary components.
\item Two worldsheets that bound $\Gamma_{1}$ and $\Gamma_{2}$.
\end{enumerate}
As in the case of confinement, the worldsheets may be self-intersecting. In the second case, the two separate worldsheets may
also overlap.  Note that worldsheets of Type 2 will become insensitive to separation in the continuum limit; the non-trivial
behavior of interest thus lies in the Type 1 worldsheets. 

We note in passing that by all the same arguments provided in Section 3 and 4, terms contributing 
to $\left< O_{\Gamma_{1} \Gamma_{2}(x)} \right>$ will have the canonical form of
background amplitudes multipled by a sum of worldsheet interaction factors that provide an effective tension.
The main difference is that surface ``entropy'', or number of Type 1 worldsheets of a given area may have a different
growth rate ($T_{0}$ in the notation used above) on account of their different topology 
With this in mind, an entirely analogous argument to that of Section 4 can be made
with the conclusion that if the inequality~(\ref{inequality}) is satisfied (with the possible exception of a
vanishing weight subset in the continuum limit), one obtains a converging series in Type 1 worldsheets of increasing area, 
the leading order term of which will be a Type 1 worldsheet corresponding straight cylinder connecting the two plaquettes.
Eventually we hope that similar analyses can be applied to arbitrary $n$-point functions, possibly leading to a
rigorous definition of the quantum field theory.
\end{appendices}


\begin{thebibliography}{99}
\bib{AA09}{article}{
    author = {Abdesselam, A},
    title = {On the Volume Conjecture for classical spin networks},
    year = {2009},
    eprint = {arXiv:0904.1734v2},
}
\bib{AS}{article}{
    author={Anishetty, R},
    author={Sharatchandra, H S},
    title={Duality transformation for non-Abelian lattice gauge theories},
    journal={Phys. Rev. Lett.},
    year = {1990},
    volume = {65},
    pages = {813--815},
}
\bib{ACSM}{article}{
    author={Anishetty, R},
    author={Cheluvarajai, S},
    author={Sharatchandra, H S},
    author={Mathur, M},
    title={Dual of a 3-dimensional pure $SU(2)$ Lattice Gauge Theory and the Ponzano-Regge Model},
    journal={Phys. Lett. B.},
    year = {1993},
    volume={314},
    pages = {387--390},
}
\bib{AGMS91}{article}{
    author = {Anishetty, R},
    author = {Gadiyar, G H},
    author = {Mathur, M},
    author = {Sharatchandra, H S},
    title = {Color invariant additive fluxes for SU(3) gauge theory},
    journal = {Phys. Lett. B},
    year = {1991},
    volume = {271}, 
    pages = {391--394},
}
\bib{stringyantonov}{article} {
    author = {Antonov, D},
    title = {String nature of confinement in (non-)abelian Gauge Theories},
    journal = {Surveys High Energ. Phys.},
    volume = {14},
    year = {2000},
    pages = {265--355},
}
\bib{A1992}{article}{
    author = {Archer, G J},
    title = {The $U_{q}$sl(3) $6-j$ symbols and state sum model},
    year = {1992},
    journal = {Physics Letters B},
    volume = {295},
    pages = {199--208},
}
\bib{BDMUV}{article} {
    author = {Burgio, C},
    author = {de Pietri, R},
    author = {Morales-Tecotl, H A},
    author = {Urrutia, L F},
    author = {Vergara, J D},
    title = {The basis of the physical Hilbert space of lattice gauge theories},
    journal = {Nucl. Phys. B},
    volume = {566},
    year = {2000},
    pages = {547--561},
}
\bib{BDMUV2}{article} {
    author = {Burgio, C},
    author = {de Pietri, R},
    author = {Morales-Tecotl, H A},
    author = {Urrutia, L F},
    author = {Vergara, J D},
    title = {The basis of the physical Hilbert space of lattice gauge theories},
    journal = {Nucl. Phys. Proc. Suppl.},
    volume = {83},
    year = {2000},
    pages = {926--928},
}
\bib{CherringtonFermions}{article}{
    author = {Cherrington, J W},
    title = {A Dual Algorithm for Non-abelian Yang-Mills coupled to Dynamical Fermions},
    year = {2008},
    journal = {Nuclear Physics B},
    volume = {794},
    pages = {195--215},
}
\bib{LAT08}{article}{
    author = {Cherrington, J W},
    title = {Recent Developments in Dual Lattice Algorithms},
    journal = {PoS LATTICE2008:050},
    eprint={arXiv:0810.0546v1},
}
\bib{Conf2}{article}{
    author = {Cherrington, J W},
    title = {A Gauge-independent mechanism for confinement and mass gap: Part II --- SU(2) in D=3},
}
\bib{CCK}{article}{
    author={Cherrington, J W},
    author={Christensen, J D},
    author={Khavkine, I},
    title={Dual computations of non-abelian {Y}ang-{M}ills on the lattice},
    journal={Physics Review D},
    year={2007},
    volume={76},
    pages={094503\ndash 094519},
    eprint={arXiv:0705.2629},
}
\bib{CC2009}{article}{
    author={Cherrington, J W},
    author={Christensen, J D},
    title={A Dual Non-abelian Yang-Mills Amplitude in Four Dimensions},
    journal={Nuclear Physics B},
    year={2009},
    volume={813},
    pages={370-382},
    eprint={arXiv:0808.3624},
}
\bib{ConradyPolyakov}{article}{
    author = {Conrady, F},
    title = {Dual representation of Polyakov loop in 3d SU(2) lattice Yang-Mills theory},
    eprint = {arXiv:0706.3422v2},
}
\bib{ConradyGluons}{article}{
    author = {Conrady, F},
    title = {Analytic derivation of dual gluons and monopoles from $SU(2)$ lattice Yang-Mills theory. 
      II. Spin foam representation},
    year = {2006},
    eprint = {arXiv:hep-th/0610237v4},
}
\bib{ConradyKhavkine}{article} {
    author = {Conrady, F},
    author = {Khavkine, I},
    title = {An exact string representation of 3d SU(2) lattice Yang--Mills theory},
    eprint = {arXiv:0706.3423v1},
}
\bib{DZ1983}{article}{
    title = {Strong Coupling and Mean Field Methods in Lattice Gauge Theories},
    author = {Drouffe, J M},
    author = {Zuber, J B},
    journal = {Physics Reports},
    volume = {102},
    year = {1983},
}
\bib{FNR06}{article}{
    author = {Freidel, L},
    author = {Noui, K},
    author = {Roche, P},
    title = {6J Symbols Duality Relations},
    year = {2007},
    eprint = {arXiv:hep-th/0604181v1},
    journal = {J. Math. Phys.},
    volume = {48},
    pages = {113512},
}
\bib{GVdW09}{article}{
    author = {Garoufalidis, S},
    author = {Van der Veen, R},
    title = {Asymptotics of Classical Spin Networks},
    year = {2009},
    eprint = {arXiv:0902.3113v1},
}
\bib{GU08}{article}{
    author = {Gurau, Razvan},
    year = {2008},
    title = {The Ponzano-Regge asymptotic of the 6j symbol: an elementary proof},
    pages = {1413-1424},
    journal = {Annales Henri Poincare},
    volume = {9},
}
\bib{itzykrough}{article} {
    author = {Itzykson, C},
    title = {Lattice Gauge Theory and Surface Roughening},
    journal = {Phys. Scr.},
    volume = {24},
    pages = {854-866},
}
\bib{IDV1}{book} {
    title = {Statistical Field Theory: Volume 1, From Brownian Motion to Renormalization and Lattice Gauge Theory},
    author = {Itzykson, C},
    author = {Drouffe, J M},
    year = {1991},
}
\bib{IDV2}{book} {
    title = {Statistical Field Theory: Volume 2, Strong Coupling, Monte Carlo Methods, Conformal
    Field theory and Random Systems},
    author = {Itzykson, C},
    author = {Drouffe, J M},
    year = {1991},
}
\bib{KIM2006}{article}{
    author = {Kim, D},
    title = {Graphical Calculus on Representations of Quantum Lie Algebras},
    journal = {Journal of knot theory and its ramifications},
    volume = {15(4)},
    year = {2006},
    pages = {453--469},
}
\bib{K97}{article}{
    title = {Spiders for Rank 2 Lie Algebras},
    author = {Kuperberg, G},
    volume = {180},
    pages = {109--151},
    year = {1997},
}
\bib{LD09}{article}{
    title = {Pushing Further the Asymptotics of the $6j$-symbol},
    author = {Levine, E},
    author = {Dupuis, M},
    year = {2009},
    eprint = {arXiv:0905.4188v1},
}
\bib{LJ09}{article}{
    author = {Littlejohn, R G},
    title = {Uniform Semiclassical Approximation for the Wigner $6j$ Symbol},
    year = {2009},
    eprint = {arXiv:0904.1734v2},
}
\bib{dsupermandelstam}{article} {
    author = {Mandelstam, S},
    journal = {Phys. Rep.},
    volume = {23C},
    pages = {245},
    year = {1976},
}
\bib{Menotti}{article} {
    author = {Menotti, P},
    author = {Onofri, E},
    author = {},
    title = {The action of $SU(N)$ lattice gauge theory in terms of the heat kernel on the group manifold},
    journal = {Nucl. Phys. B},
    volume = {190},
    pages = {288--300},
    year = {1981},
}
\bib{M1980}{article}{
    title = {High Temperature Expansions for the Free Energy of Vortices and the String Tension in Lattice Gauge Theories},
    author = {Munster, G},
    journal = {Nuclear Physics B},
    volume = {180},
    pages = {23-60},
    year = {1981},
}
\bib{MandM}{book}{
    author = {Montvay, I},
    author = {Munster, G},
    title = {Quantum Fields on a Lattice},
    year = {1994},
    publisher = {Cambridge University Press},
    place = {Cambridge},
}
\bib{stringynambu}{article}{ 
    author = {Nambu, Y},
    journal = {Phys. Lett. B},
    volume = {80},
    year = {1979},
    pages = {372},
}
\bib{dsuperparisi}{article} {
    author = {Parisi, G},
    journal = {Phys. Rev.},
    volume = {D11},
    pages = {971},
    year = {1975},
}
\bib{OecklDGT}{book}{
    author={Oeckl, R},
    title={Discrete Gauge Theory: From Lattices to {TQFT}},
    year={2005},
    publisher={Imperial College Press},
    place={London, UK},
}
\bib{OecklPfeiffer}{article}{
    author={Oeckl, R},
    author={Pfeiffer, H},
    title={The dual of pure non-Abelian lattice gauge theory as a spin foam model},
    journal={Nuclear Physics B},
    year={2001},
    volume={598},
    pages={400\ndash 426},
}
\bib{Osterwalder}{article} {
    author = {Osterwalder, K},
    author = {Seiler, E},
    title  = {Gauge Field Theories on a Lattice},
    year = {1978}, 
    journal = {Annals of Physics},
    volume = {110}, 
    pages = {440--471},
}
\bib{cvortexthooft}{article} {
    author = {'t Hooft, G},
    year = {1978},
    journal = {Nuclear Physics B},
    volume = {1},
}
\bib{dsuperthooft}{article} {
    author = {'t Hooft, G},
    book = { 
       title = {High Energy Physics},
       year = {1975},
       editor = {Zichichi, A},
       place = {Palermo},
    },
}
\end{thebibliography}
\end{document}